\documentclass[11pt,fleqn]{elsart}

\usepackage{amsmath}
\usepackage{amssymb}
\usepackage{theorem}
\usepackage{xspace}
\usepackage{natbib}
\usepackage{epsfig}

\def\pth#1{\left(#1\right)}
\def\acc#1{\left\{#1\right\}}
\def\cro#1{\left[#1\right]}

\def\eR{I\!\!R}
\def\eE{I\!\!E}
\def\eP{I\!\!P}
\def\e1{1\!\!1}

\def\el{{\lambda}'}
\def\ef{\mathbf{\overset{.}{f}}}
\def\eff{\mathbf{\overset{..}{f}}}
\def\eg{\mathbf{g}}
\def\ex{\mathbf{x}}
\def\eX{\mathbf{X}}

\def\ebo{\textrm{\mathversion{bold}$\mathbf{\beta^0}$\mathversion{normal}}}
\def\eb{\textrm{\mathversion{bold}$\mathbf{\beta}$\mathversion{normal}}}  
\def\el{\textrm{\mathversion{bold}$\mathbf{\lambda}$\mathversion{normal}}}  

\newtheorem{remark}{Remark}
\newtheorem{theorem}{Theorem}[section]
\newtheorem{corollary}{Corollary}[section]
\newtheorem{lemma}{Lemma}[section]

\newcommand{\R}{\mathbb{R}}

\def\argmin{\mathop{\mathrm{arg\,min}}} 
\begin{document}
\begin{frontmatter}

\title{Random nonlinear model with missing responses }

\author{Gabriela CIUPERCA}\footnote{Tel:(33)0472431690; Fax:(33)0472431687\\ {\it E-mail address: Gabriela.Ciuperca@univ-lyon1.fr} }

\address{Universit\'e de Lyon, Universit\'e Lyon 1, 
UMR 5208, blvd du 11 novembre 1918,
F - 69622 Villeurbanne Cedex, France}

\begin{abstract}
A nonlinear model with response variable missing at random  is studied. In order to improve the coverage accuracy, the empirical likelihood ratio (EL) method is considered. The asymptotic distribution  of EL statistic and also of its approximation is $\chi^2$ if the parameters are estimated using least squares(LS) or least absolute deviation(LAD) method on complete data. When the  response are reconstituted using a semiparametric method, the empirical log-likelihood associated on imputed data is also asymptotically $\chi^2$.  The Wilk's theorem  for EL for parameter on response  variable is also satisfied. It is shown via Monte Carlo simulations that the EL methods outperform the normal approximation based method in terms of coverage probability up to and including on the reconstituted data. The advantages of the proposed method are exemplified on the real data. \\
{\it keywords:} random nonlinear model; response missing at random; empirical likelihood; semi-parametric estimation;
\end{abstract}
\end{frontmatter}


\section{Introduction}
It is very common in practice, for a model, to measure the regressors(covariates) but for varied reasons it is sometimes impossible to have all values of the response variable. The most widely used idea is to remove of model the observations with missing data. An  alternative solution is to consider empirical likelihood(EL) method which is a powerful nonparametric method for constructing confidence regions of parameters.\\
To our knowledge, previous theoretical and numerical investigations in literature have focused for  model with missing response only in the linear case. The  EL method, proposed by \cite{Owen:90} does not need the asymptotic variance of estimator and it outperforms the normal approximation method in term of coverage probability for linear models. \cite{Wang:Rao:02} develops EL inferences for the mean of a response variable under regression imputation of missing responses for a linear regression model and random covariates. \cite{Qin:Li:Lei:09} construct an EL statistic on parameter when regressors are deterministic and \cite{Xue:09} if regressors are random, based on least squares(LS) method for linear model $Y=\eX^t \eb +\varepsilon$. \cite{Sun:Wang:09} consider the general linear model $Y=\textbf{H}(\eX)^t\eb+\varepsilon$ with $\textbf{H}(\ex)$ a known vector function and investigate a hypothesis test on the response variable. These last three papers impose the condition that the conditional expectation of error $\varepsilon$ with respect to covariate $\eX$ is zero. If this hypothesis is not satisfied, in order to reconstitute the response variable  the least absolute deviations(LAD) method can be used. One advantage of least absolute deviations estimation is that it does not require any moment condition on the errors to obtain asymptotic normality. \\
It is well known also that one outlier may cause a large error in a least squares estimator. This occurs in the case of fatter tail distributions of the error term. On the other hand, as \cite{Bai:98} and \cite{Kim:Choi:95} indicate it, for heavy tailed distributions the LAD estimator is more efficient than LS estimator.  \\
Concerning the LAD estimator in a complete nonlinear model we can refer to following papers: \cite{Oberhofer:82} shows conditions for its consistency, \cite{Weiss:91} proves that this estimator is consistent and asymptotically normal in a dynamic nonlinear model with neither independent nor identically distributed errors. \cite{Ciuperca:09} gives the convergence rate $v_n$, where $(v_n)$ is a monotone positive sequence such that $v_n \rightarrow 0$ and $n v_n^2 \rightarrow \infty$ for $n \rightarrow \infty$.
It is well known that confidence regions based on the asymptotic normality could encounter large coverage errors in small sizes or if the error distribution has outliers. The LAD technique was already used in censored median linear regression model with missing data: see e.g. \cite{Zhao:Chen:08}. \\
For other relevant papers (not exhaustive list) on EL method for missing data in a linear models see \cite{Wang:Sun:07}, \cite{Sun:Wang:Dai:09}, \cite{Wang:Linton:Hardle:04}, \cite{Yang:Xue:Cheng:09}, \cite{Liang:Qin:Zhang:Ruppert:09}.\\
In this paper, for a nonlinear random model $Y=f(\eX;\eb)+\varepsilon$ with missing responses, some empirical likelihood ratios are constructed by using complete-case or imputed values. The nonparametric version of Wilks' theorem is proved for two cases: the parameters are estimated by  least squares and least absolute deviations on complete data. The limiting distribution of EL statistic is $\chi^2$, results which can be used to construct confidence region on parameter. In order to complete data, the value imputed of  missing value of variable response is obtained by generalising  \cite{Xue:09} idea for linear model using a semiparametric technique: the parameters regression are estimated by LS method and missing probability by a nonparametric method. We show that the empirical log-likelihood ratio on parameter based on the improved  data is asymptotically chi-squared. The numerical simulations proves that EL methods outperforms the normal approximation  method in terms of coverage probability up to and including on the reconstituted data. If the distribution of the errors presents outliers, the LAD method gives generally best results that LS method on coverage probability and on parameter estimators efficiency. In addition to that, if the expectation of error does not exist, as it is the case for Cauchy distribution, the normal approximation of LS estimator can not be satisfied.  On the real data we obtain also that our semiparametric method gives more precise results to reconstitute the response variable that classic parametric LS method.  \\
The rest of this paper is organised as follows. In Section 2 we introduce model, assumptions and some notations. In Section 3 the Wilks' theorem for EL statistic and also for its approximation is given, when the parameters are estimated using LS or LAD method on complete data. The LS case is developed in Section 4, a reconstituted value for the response variable is introduced and asymptotic distribution of EL for response variable is obtained. Section 5 illustrates by simulation results that EL methods for nonlinear random model outperform the normal method and give very competitive coverage probabilities. An application to the real data is presented in Section 6. Finally, Section 7 contains the proofs of the lemmas and of the theorems. 

\section{Notations and model}
Let us consider following random nonlinear model:
\begin{equation}
 \label{e1}
Y_i=f(\eX_i,\eb)+\varepsilon_i, \qquad i=1,\cdots,n,
\end{equation}
where $(\varepsilon_i,\eX_i)$ is a sequence of continuous independent random vectors with the same joint distribution as $(\varepsilon,\eX)$ and $\eb$ is a $d \times 1$ vector of unknown regression parameters. More precisely, $\varepsilon$ is a random variable and $\eX$ a $p \times 1$ random vector of covariates.  Let $\ebo$ denote the true (unknown) of the parameter $\eb$.\\
With regard to the random variable $\varepsilon$ we make one of the suppositions:\\
{\bf (H1)} $\eE[\varepsilon_i|\eX_i]=0$ and $\eE[\varepsilon_i^2|\eX_i] < \infty$, $\forall i=1, \cdots,n$.\\
{\bf (H1bis)} $\eE[sign(\varepsilon_i)|\eX_i]=0$ and $\varepsilon_i$ have continuous density $e(t)$ satisfying $e(0)>0$.\\
These conditions are essential for the consistency and asymptotic normality of the LS, respectively LAD estimator.\\

In following we use notation $\ef(\ex,\eb)=\partial f(\ex,\eb)/\partial \eb$, $\eff(\ex, \eb)=\partial^2 f(\ex,\eb)/\partial \eb^2$. For a vector, let us denote $\|.\|$ the Euclidean norm.\\
Regression function $f:\Upsilon \times \Gamma \rightarrow \R$, with $\Upsilon \subseteq \R^p$, $\Gamma \subseteq \R^d$ and random vector $\eX$ satisfy the conditions:\\
\textbf{(H2)} for all $\ex \in \Upsilon$ and for $\eb \in \Gamma$, the function $f(\ex,\eb)$ is twice differentiable in $\eb$ and continuous on $\Upsilon$.\\
\textbf{(H3)} $\eP[\|\eX\| >M_n]=o(n^{-1/2})$ for any positive sequence $M_n \rightarrow \infty$ as $n \rightarrow \infty$.\\
\textbf{(H4)}  $ \| \ef(\ex,\eb) \|$, $\| \eff(\ex,\eb) \| $ are  bounded for any $ \ex \in \Upsilon$ and  $\eb$ in a neighborhood of $\ebo$. \\
Sets $\Upsilon$ and $\Gamma$ are compacts. This, assumptions (H2) and (H4) are commonly used in nonlinear modelisation and are necessary for the consistency and for the asymptotic normality of the LS or LAD estimator. Assume also that the model is identifiable: if $f(\eX; \eb)=f(\eX; \eb^*)$ with probability one, then $\eb=\eb^*.$\\

For model (\ref{e1}), all the $\eX_i$'s are observed, in exchange response variable $Y_i$ can be missing. Let be the sequence of random variables $(\delta_i)_{1 \leq i \leq n}$ defined by: $\delta_i=0$ if $Y_i$ is missing and $\delta_i=1$ if $Y_i$ is observed. We suppose that $Y_i$ is missing at random(MAR): $\eP[\delta_i=1|\eX_i,Y_i]=\eP[\delta_i=1|\eX_i]$, for all $i=1,\cdots,n$. Consider the selection probability function: $\pi(\ex)=\eP[\delta=1 |\eX=\ex]>0$, $\forall \ex \in \Upsilon$.  
The supposition $\pi(\ex)>0$ is a common assumption in the literature. The parameter $\eb$ is estimated on the completely observed data by two methods: least squares(LS) method:
\begin{equation}
 \label{e2}
\hat \eb_{n,LS}=\argmin_\eb \sum^n_{i=1} \delta_i [Y_i-f(\eX_i,\eb)]^2
\end{equation}
and least absolute deviations(LAD) method:
 \begin{equation}
 \label{e3}
\hat \eb_{n,LAD}=\argmin_\eb \sum^n_{i=1} \delta_i |Y_i-f(\eX_i,\eb)|.
\end{equation}
To build the EL  statistic, let us consider following functions, for $i=1, \cdots,n$: 
\begin{equation*}
\textbf{g}_{i,LS}(\eb)=\delta_i [Y_i-f(\eX_i,\eb)] \ef(\eX_i,\eb),
\end{equation*}
\begin{equation*}
 \eg_{i,LAD}(\eb)=\delta_i sign(Y_i-f(\eX_i,\eb)) \ef(\eX_i,\eb)
\end{equation*}
and let be also:
\begin{equation*}
 \eg(\eb)=\frac{1}{n} \sum^n_{i=1} \eg_i(\eb),
\end{equation*}
with $\eg_i(\eb)$ either $\eg_{i,LS}(\eb)$ or $\eg_{i,LAD}(\eb)$. The two estimators are a solution of the system of equations: $\eg(\eb)=\textbf{0}$. Under (H1), respectively (H1bis), we have: $\eE[\eg_i(\eb)]=\textbf{0}$ for $\eg_i=\eg_{i,LS}$, respectively $\eg_{i,LAD}$. The joint distribution of the errors $\varepsilon_i$ and of $\eX_i$ is unknown, but conditional mean, respectively median, of $\varepsilon_i$ is zero.

The empirical likelihood for $\eb$ with complete-case data can be defined as (see \cite{Owen:90}):
\begin{equation}
 \label{e4}
L_n(\eb)=\sup_{(p_1, \cdots, p_n) \in (0,1)^n} \acc{ \prod^n_{i=1} p_i; \quad \sum^n_{i=1} p_i=1, \sum^n_{i=1} p_i \eg_i(\eb)=\textbf{0}}.
\end{equation}
Without constraint $ \sum^n_{i=1} p_i \eg_i(\eb)=\textbf{0}$, the maximum of $\prod^n_{i=1} p_i$ is attained for $p_i=n^{-1}$. Then, the profile empirical likelihood ratio for $\eb$ with complete-case has the form:
\begin{equation}
 \label{e5}
R_n(\eb)=\sup_{(p_1, \cdots, p_n) \in (0,1)^n} \acc{ \prod^n_{i=1} np_i; \quad \sum^n_{i=1} p_i=1, \sum^n_{i=1} p_i \eg_i(\eb)=\textbf{0}}.
\end{equation}
The empirical log-likelihood ratio statistic evaluated at $\eb$ is:
\begin{equation}
 \label{e6}
l_n(\eb)=-2 \sup_{(p_1, \cdots, p_n) \in (0,1)^n} \acc{ \sum^n_{i=1} \log(np_i); \quad \sum^n_{i=1} p_i=1, \sum^n_{i=1} p_i \eg_i(\eb)=\textbf{0}}.
\end{equation}
Let $\el \in \eR^d$ be, a Lagrange multiplier. Then, $R_n(\eb)$  is maximised for $
p_i=n^{-1} [1+\el^t \eg_i(\eb)]^{-1}
$, 
where $\el$ satisfies the equation:
\begin{equation}
 \label{e8}
\textbf{0}=\frac{1}{n} \sum^n_{i=1} \frac{\eg_i(\eb)}{1+\el^t \eg_i(\eb)}.
\end{equation}
Thus, $l_n(\eb)$ can be written:
\begin{equation}
 \label{e7}
l_n(\eb)=2\sum^n_{i=1} \log (1+\el^t \eg_i(\eb)).
\end{equation}
In order to study the asymptotic properties of the EL statistic given by (\ref{e7}), let us consider following matrix: $
\textbf{A}=\eE \cro{\pi(\eX) \ef(\eX,\ebo)\ef^t(\eX,\ebo) }
$ and under assumption (H1): $
\textbf{B}=\eE \cro{ \varepsilon^2(\eX)\pi(\eX) \ef(\eX,\ebo)\ef^t(\eX,\ebo) }
$. 
On suppose for the first matrix:\\
{\bf (HA)} $\textbf{A}$ is a positive definite matrix.\\
Let us notice that $\textbf{A}$ is Fisher information matrix on complete data.
\section{Asymptotic properties}
We give first a classical result for a MAR model, lemma that turn out to be useful in the proof of the main results. 
\begin{lemma}
 \label{lemma2}
Under assumptions (H1), respectively (H1bis) and (H2), (H4), (HA), we have:\\
(i) $\frac{1}{\sqrt n} \sum^n_{i=1} \eg_i(\ebo) \overset{{\cal L}} {\underset{n \rightarrow \infty}{\longrightarrow}} {\cal N}_d(\textbf{0}, \textbf{V}) $,\\
(ii) $\frac{1}{n}\sum^n_{i=1} \eg_i(\ebo) \eg_i^t(\ebo) \overset{{\eP}} {\underset{n \rightarrow \infty}{\longrightarrow}} \textbf{V}$,\\
(iii) $\max_{1 \leq i \leq n} \|\eg_i(\ebo) \| =o_{\eP}(n^{1/2})  $,\\
with $\textbf{V}=\textbf{B}$ for LS and $\textbf{V}=\textbf{\textbf{A}}$ for LAD method.
\end{lemma}
Let us consider following matrix: 
\begin{equation}
 \label{Bn}
\textbf{B}_n=n^{-1} \sum^n_{i=1} \eg_i(\ebo) \eg_i^t(\ebo).
\end{equation} 
Following theorem gives the asymptotic distribution of the empirical log-likelihood statistic (\ref{e7}), evaluated at the true value.  \begin{theorem}
 \label{theorem1}
Under assumptions (H1), respectively (H1bis), (H2), (H4), (HA) $l_n(\ebo)\overset{{\cal L}} {\underset{n \rightarrow \infty}{\longrightarrow}} \chi^2(d)$.
\end{theorem}
We can use Theorem \ref{theorem1} to get approximate confidence region for $\eb$ or for testing the hypothesis: $H_0:\eb=\ebo$. Since, by the proof of Theorem \ref{theorem1} (see Appendix) for the EL statistic we have: \\
$l_n(\ebo)=\textbf{B}_n^{-1} \pth{\frac{1}{\sqrt n} \sum^n_{i=1}  \eg_i^t(\ebo) } \pth{\frac{1}{\sqrt n} \sum^n_{i=1}  \eg_i(\ebo) }(1+o_{\eP}(1))$, in order to calculate numerically $l_n(\ebo)$ we  can use the approximation:
\begin{equation}
 \label{e19}
l_n^*(\ebo)=\textbf{B}_n^{-1} \pth{\frac{1}{\sqrt n} \sum^n_{i=1}  \eg_i^t(\ebo) } \pth{\frac{1}{\sqrt n} \sum^n_{i=1}  \eg_i(\ebo) }.
\end{equation}
We state this as a corollary.
\begin{corollary}
\label{corollaire1}
 Under the same assumptions as in Theorem \ref{theorem1}, the asymptotic distribution of $l_n^*(\ebo)$ is $\chi^2(d)$. 
\end{corollary}

Thus, an asymptotic $(1-\alpha)$ confidence region for $\eb$, based on EL statistic on complete data is: $\{ \eb; l_n^*(\eb) \leq c_{1-\alpha;d} \}$ where $c_{1-\alpha;d}$ is the $(1-\alpha)$ quantile of the chi-squared distribution with degrees of freedom $d$. It is very interesting to note that to construct the confidence region for $\eb$ it is not necessary to calculate the Lagrange multiplier which intervenes in (\ref{e7}), observed data are enough.\\
Asymptotic normality of LS and LAD estimators calculated on complete data is given by the following result.
\begin{theorem}
 \label{theorem2}
(i) Under assumption (H1), (H2), (H4), (HA) we have:\\
 ${\sqrt n} (\hat \eb_{n,LS}- \ebo) \overset{{\cal L}} {\underset{n \rightarrow \infty}{\longrightarrow}} {\cal N}_d(\textbf{0}, \textbf{A}^{-1}\textbf{B} \textbf{A}^{-1})$.\\
(ii) Under assumption (H1bis), (H2), (H4), (HA) we have:\\
  $2 {\sqrt n} e(0) \textbf{A}^{1/2}  (\hat \eb_{n,LAD}- \ebo) \overset{{\cal L}} {\underset{n \rightarrow \infty}{\longrightarrow}} {\cal N}_d(\textbf{0}, \textbf{I}_d)$.
\end{theorem}
These theorem allows to give the normal approximation based confidence region, expression which will be specified in Section 5. Then, on complete data, we have the choice between four statistics ($l_n^*(\ebo)$ for LS, $l_n^*(\ebo)$ for LAD, $\hat \eb_{n,LS}$, $\hat \eb_{n,LAD}$) to  test hypotheses or to build the asymptotic  confidence region of model parameter $\eb$. We see in Sections 5 and 6, by simulations and a model on real data, that approximated EL statistics are sharply superior to normal approximation given by Theorem \ref{theorem2}. If error distribution presents outliers then $l^*_n(\eb)$ for LAD method is recommended, otherwise it is better to consider $l^*_n(\eb)$ for LS method. This last one will more be developed in the following section. The missing probabilities $\pi(\eX_i)$ are estimated by a nonparametric method, this is going to allow to reconstitute the missing responses. On the observed and the reconstituted observations one defined a new EL statistic, which also satisfies a Wilk's theorem. Besides, numerically, it gives very competitive results (see Sections 5 and 6).  
\section{Special case of LS estimator}
Following e.g. \cite{Xue:09}, \cite{Qin:Li:Lei:09} for linear model, we shall introduce the  forecast of $Y_i$, constructed by using LS estimator for parameter $\eb$ and  a nonparametric estimator for probability $\pi(\eX_i)$:
\begin{equation}
\label{yy}
 Y_{n,i}=\frac{\delta_i}{\hat \pi(\eX_i)}Y_i+\pth{1-\frac{\delta_i}{\hat \pi(\eX_i)}} f(\eX_i;\hat \eb_{n,LS}), \qquad i=1, \cdots, n,
\end{equation}
with $\hat \pi(\eX_i)$ a nonlinear estimator for $\pi(\eX_i)$, as in the linear regression \cite{Xue:09}:
\begin{equation*}
 \hat \pi(\eX_i)=\frac{\sum^n_{i=1} \delta_i K_h((\eX_i-\ex)/h)}{\max\{ 1, \sum^n_{i=1}  K_h((\eX_i-\ex)/h)\}},
\end{equation*}
where $h=h_n$ is a positive sequence tending to 0 as $n \rightarrow \infty $ and $K_h$ is a kernel function defined in $\R^d$. The bandwidth $h_n$ satisfies:\\
\textbf{(H5)} $n h_n^{2d} M_n^{-2d}\rightarrow \infty$ and $n h_n^{4 \max \{ 2,d-1 \}}\rightarrow 0$, as $n \rightarrow \infty $, with the sequence $M_n$ given in assumption (H3).\\
The kernel function $K$, satisfies the classical condition (imposed also for the linear model using the LS method of \cite{Xue:09}):\\
\textbf{(H6)} there exist positive constants $C_1$, $C_2$ and $\rho$ such that: $
C_1 \e1_{\|u\| \leq \rho} \leq K(u) \leq C_2 \e1_{\|u\| \leq \rho}$. \\
Concerning the selection probability function $\pi(\eX)$ let us make following regularity hypothesis necessary in the study of its nonparametric estimator. \\
\textbf{(H7)} $p(\ex)$ has bounded partial derivatives up to order $\max(2,d-1)$ almost everywhere.\\
Conditions (H5)-(H7) are usual assumptions for convergent rates of kernel estimating method.

Let us denote $\theta^0=\eE[Y]$ the mean of $Y$ and  $\sigma^2(\ex)=\eE[\varepsilon^2 |\eX=\ex]$ error variance.\\
Following lemma gives the asymptotic normality for the sequence $(Y_{n,i})$ and other two similar results of Lemma \ref{lemma2}.
\begin{lemma}
 \label{lemma3} Under assumptions (H1)-(H6) and if $Var[f(\eX,\ebo)] < \infty$ we have:\\
(i) $\frac{1}{\sqrt n} \sum^n_{i=1} (Y_{n,i}-\theta^0)\overset{{\cal L}} {\underset{n \rightarrow \infty}{\longrightarrow}}{\cal N}(0, W)$,\\
(ii) $\frac{1}{n} \sum^n_{i=1} (Y_{n,i}-\theta^0)^2\overset{{\eP}} {\underset{n \rightarrow \infty}{\longrightarrow}} W$,\\
(iii) $\max_{1 \leq i \leq n} | Y_{n,i}| =o_{\eP}(n^{1/2})$,\\
with $W=Var[f(\eX,\ebo)]+\eE \cro{\frac{\sigma^2(\eX)}{\pi(\eX)}}$.
\end{lemma}

Using similar arguments as for Theorem \ref{theorem1} we obtain the  following result. We hence omit its proof.
\begin{theorem}
 \label{theorem3}
Suppose that  assumptions (H1)-(H6) hold, then for empirical log-likelihood for $\theta^0$:
\begin{equation*}
l_{n,Y}(\theta^0)=-2 \sup_{(p_1, \cdots, p_n) \in (0,1)^n} \acc{ \sum^n_{i=1} \log(np_i); \quad \sum^n_{i=1} p_i=1, \sum^n_{i=1} p_i Y_{n,i}=\theta^0},
\end{equation*}
we have: $l_{n,Y}(\theta^0) \overset{{\cal L}} {\underset{n \rightarrow \infty}{\longrightarrow}} \chi^2(1)$.
\end{theorem}
This result can be used to make test of hypothesis or to construct asymptotic confidence region for the response variable.
\cite{Xue:09b} constructs a weight-corrected empirical log-likelihood ratio for $\theta^0$ which is also asymptotically chi-squared.\\
Let be now following  functions constructed using the reconstituted response:
\begin{equation*}
\eg_{n,i}(\eb)=[Y_{n,i}-f(\eX_i;\eb)] \ef(\eX_i;\eb), \qquad i=1, \cdots,n.
\end{equation*}
Consider also the empirical log-likelihood associated at $\eg_{n,i}(\eb)$:
\begin{equation*}
\hat l_n(\eb)=-2 \sup_{(p_1, \cdots, p_n) \in (0,1)^n} \acc{ \sum^n_{i=1} \log(np_i); \quad \sum^n_{i=1} p_i=1, \sum^n_{i=1} p_i \eg_{n,i}(\eb)=0}.
\end{equation*}
Then, the equivalent of (\ref{e7}) is:
\begin{equation}
 \label{e7bis}
\hat l_n(\eb)=2\sum^n_{i=1} \log (1 +\el^t \eg_{n,i}(\eb)).
\end{equation}
Consider  following lemma needed for  Theorem \ref{theorem5}.
\begin{lemma}
 \label{lemma4}
Under assumptions (H1)-(H7), we have:\\
(i) $ n^{-1/2} \sum^n_{i=1} \eg_{n,i}(\ebo) \overset{{\cal L}} {\underset{n \rightarrow \infty}{\longrightarrow}}{\cal N}_d \pth{\textbf{0}, \eE \cro{\pi(\eX)^{-1} \varepsilon^2(\eX) \ef(\eX;\ebo) \ef^t(\eX;\ebo)}}$.\\
(ii) $n^{-1}\sum^n_{i=1} \eg_{n,i}(\ebo) \eg_{n,i}^t(\ebo) \overset{{\eP}} {\underset{n \rightarrow \infty}{\longrightarrow}} \eE \cro{\pi(\eX)^{-1} \varepsilon^2(\eX) \ef(\eX;\ebo) \ef^t(\eX;\ebo)}$.\\
(iii) $\max_{1 \leq i \leq n} \|\eg_{n,i}(\ebo) \| =o_{\eP}(n^{1/2})  $.
\end{lemma}

Following result shows that the empirical log-likelihood ratio on $\ebo$ based on the reconstituted data converges to towards $\chi^2(d)$. This theorem shows in a similar way as Theorem \ref{theorem1}, then the proof will be omitted.
\begin{theorem}
 \label{theorem5}
Under assumptions (H1)-(H7) we have: $\hat l_n(\ebo)\overset{{\cal L}} {\underset{n \rightarrow \infty}{\longrightarrow}} \chi^2(d)$. 
\end{theorem}
From Theorem \ref{theorem5}, one can construct an asymptotic  $(1-\alpha)$-level confidence region for $\eb$ using all available  values for $\eX$ and the reconstituted values for $Y$. 

\begin{corollary}
In a similar way in the complete data, Corollary \ref{corollaire1}, the statistic $\hat l_n(\eb)$ may be approximated by:
\begin{equation}
 \label{e20}
{\hat l_n^*(\ebo)}  ={\textbf{ B}_n^*}^{-1} \pth{\frac{1}{\sqrt n} \sum^n_{i=1}  \eg_{n,i}^t(\ebo) } \pth{\frac{1}{\sqrt n} \sum^n_{i=1}  \eg_{n,i}(\ebo) },
\end{equation}
with $\textbf{ B}_n^*=n^{-1} \sum^n_{i=1}\eg_{n,i}(\ebo) \eg^t_{n,i}(\ebo)$. The asymptotic distribution of $\hat l_n^*(\ebo)$ is $\chi^2(d)$.
\end{corollary}
This implies that for testing hypothesis $H_0:\eb=\ebo$ we can use the statistic $\hat l_n^*(\ebo)$ with asymptotic reject region $\{(Y_i,\eX_i,\delta_i)_{1 \leq i \leq n} ; \hat l_n^*(\ebo) > c_{1-\alpha;d} \}$ where $c_{1-\alpha;d}$ is the $(1-\alpha)$ quantile of the chi-squared distribution with degrees of freedom $d$.
\begin{remark}
 Since the convergence rate of $\hat \eb_{n,LAD}$ to $\ebo$ can be slower than $n^{-1/2}$ (see \cite{Ciuperca:09}), then Lemma \ref{lemma4} can not be true and the analogue of the Theorem \ref{theorem5} cannot be consider for LAD estimator.
\end{remark}
We can minimise $l_n(\ebo)$ and we obtain another estimator $\tilde \eb_{n}$ of $\ebo$, called the maximum empirical likelihood estimator (MELE). Using the same arguments as used in the proof of Theorem 1 in  \cite{Qin:Lawless:94}, we obtain:
\begin{theorem}
 \label{theorem4}
Under assumptions (H1), (H2), (H4), (HA) and:
\begin{itemize}
 \item  $\eE[\eg_{i,LS}(\ebo)\eg^t_{i,LS}(\ebo)]$ is positive definite,
 \item $\partial \eg_{i,LS}(\eb)/ \partial \eb $ is continuous in a neighborhood of the true value $\ebo$,    $\|\partial \eg_{i,LS}(\eb)/ \partial \eb \|$ and $\|\eg_{i,LS}(\eb) \|^3$ are bounded by some integrable function in this neighborhood,
\item the rank of $\eE[\partial \eg_{i,LS}(\ebo)/ \partial \eb ]$ is $d$,
\item $\partial^2 \eg_{i,LS}(\eb)/ \partial \eb \eb^t$ is continuous in a neighborhood of the true value $\ebo$, $\| \partial^2 \eg_{i,LS}(\eb)/ \partial \eb \eb^t \|$ is bounded by some integrable function in this neighborhood
\end{itemize}
then\\
(i) convergence rate of $\tilde \eb_{n}$ is $n^{-1/3}$: $ \| \tilde \eb_{n} - \ebo \| =O_{\eP}(n^{-1/3})$.\\
(ii) ${\sqrt n} (\tilde \eb_n-\ebo)  \overset{{\cal L}} {\underset{n \rightarrow \infty}{\longrightarrow}}{\cal N}_d \pth{\textbf{0}, \textbf{V}_{MELE}}$, with $\textbf{V}_{MELE}=\textbf{A}^{-1} \textbf{B}_{MELE} \textbf{A}^{-1}$, $\textbf{B}_{MELE}=\eE[\pi(\eX) \varepsilon^2(\eX)\eff(\eX;\ebo)\eff^t(\eX;\ebo)]-\eE[\pi(\eX) \ef(\eX;\ebo) \ef^t(\eX;\ebo) \ef(\eX;\ebo) \ef^t(\eX;\ebo)]$.
\end{theorem}
\section{Simulation study}
In this section we use Monte Carlo simulation to compare empirical likelihood method with normal method. 
For nominal confidence level $1-\alpha=0.95$, using the simulated samples, we evaluated the coverage probabilities (CP)  of the confidence regions (CR) given by:\\
- approximated  empirical log-likelihood method on the completely observed data (Theorem \ref{theorem1}): $CR_{LS}=\acc{\eb; l^*_n(\eb) \leq  c_{1-\alpha;d}}$, $CR_{LAD}=\acc{\eb; l^*_n(\eb) \leq  c_{1-\alpha;d}}$ by LS or LAD method, respectively, where $c_{1-\alpha;d}$ is the $(1-\alpha)$ quantile of the standard chi-square distribution with $d$-degrees of freedom and $ l^*_n(\eb)$ given by (\ref{e19});\\
- approximated empirical log-likelihood $\hat l^*_n(\eb)$, given by (\ref{e20}), associated at $\eg_{n,i}(\eb)$ on the reconstituted data: ${\hat {CR}_{LS}}=\acc{\eb ; \hat l^*_n(\eb) \leq  c_{1-\alpha;d}}$;\\ 
- normal method, based on Theorem \ref{theorem2}: \\
 $NCR_{LS}=\acc{\eb; n (\hat \eb_{n,LS}-\ebo)^t \textbf{A}\textbf{B}^{-1} \textbf{A}(\hat \eb_{n,LS}-\ebo) \leq c_{1-\alpha;d}}$ and\\ $NCR_{LAD}=\acc{\eb; 4 n e(0) (\hat \eb_{n,LAD}-\ebo)^t \textbf{A}(\hat \eb_{n,LAD}-\ebo) \leq c_{1-\alpha;d}}$.\\

Throughout this section, the kernel function is taken as the Epanechnikov kernel $K(x)=0.75(1-x^2)\e1_{|x| \leq 1}$, the bandwidth sequence $h_n=n^{-1/7}$ which satisfies assumption (H5). We generate $M=2000$ Monte Carlo random samples of size $n$ for $X \sim {\cal N}(1,1)$ and  for  errors either $ {\cal N}(0,\sigma^2)$ either Laplace ${\cal L}(0,\sigma^2)$ or Cauchy ${\cal C}(0,\sigma^2)$, with $\sigma=1,2$. We consider  following two cases of response probability under the MAR assumption:\\
\textit{a)} $\pi(x)=0.8+0.2|x-1|$ if $|x-1| \leq 1$ and 0.95 elsewhere (similar the linear case of \cite{Xue:09}). The average missing rate is 0.91 and empirical mean of $\hat \pi(X_i)$ is 0.905 for $X \sim {\cal N}(1,1)$.\\
\textit{b)} $\pi(x)=0.8$ for all $x$.\\
Case a) unlike to case b) excludes most of the  error outliers. \\
The coverage probability are estimated by the frequency of the true values $\ebo$ falling into the confidence intervals in M=2000 simulations. \\
We denote by $CP_{LS}$, $CP_{LAD}$, ${\hat {CP}_{LS}}$, $NCP_{LS}$, $NCP_{LAD}$ the coverage probabilities corresponding to the confidence regions $CR_{LS}$, $CR_{LAD}$, ${\hat {CR}_{LS}}$, $NCR_{LS}$, $NCR_{LAD}$ respectively.\\
All simulations, calculations of estimations and statistical computations were performed using R language. To calculate the LS estimations on  nonlinear model we used {\it nls} function of package {\it stats} and on linear model {\it lm} function of package {\it base}. To calculate the LAD estimations function {\it nlrq} of package {\it quantreg} was used if $f$ is nonlinear and  {\it rq} function of the same package for linear model. We used also {\it VGAM} package for random generation of the Laplace distribution by {\it rlaplace} function. To generate Normal and Cauchy distribution the functions {\it rnorm} respectively {\it rcauchy} of {\it stats} package are used. 
   
\subsection{Nonlinear model}
\begin{figure}[h!]
\begin{minipage}[b]{0.53\linewidth}
\includegraphics[scale=0.48]{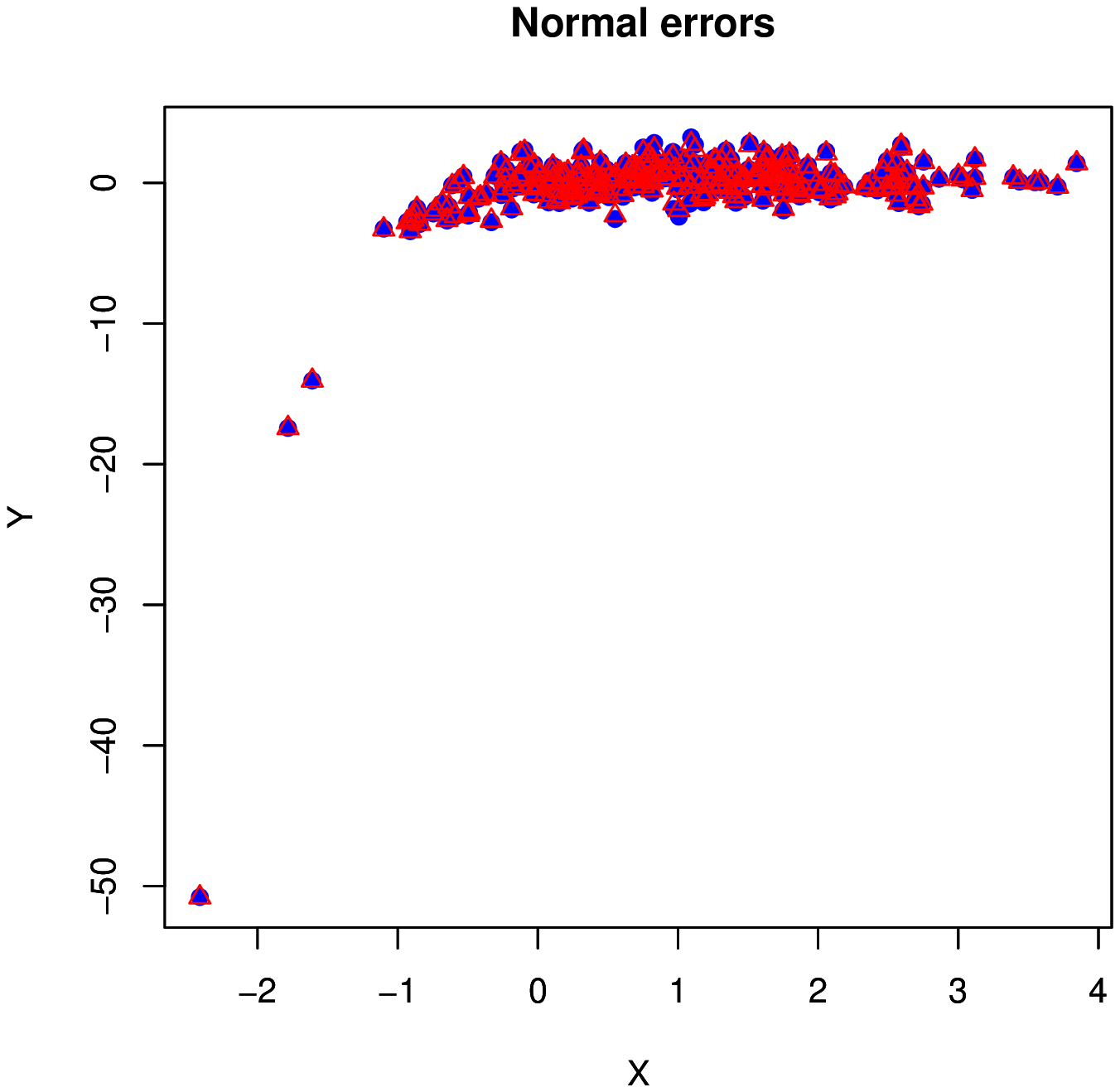} 
\caption{\it Response variable $Y$ function regressor $X$: ``solide circle'' for reconstituted values $Y_{n,i}$, ``triangle'' for true complete values $Y_i$. Normal error $\varepsilon \sim {\cal N}(0,1)$, n=300.}
\label{Figure 1}
\end{minipage}
\begin{minipage}[b]{0.53\linewidth}
\includegraphics[scale=0.48]{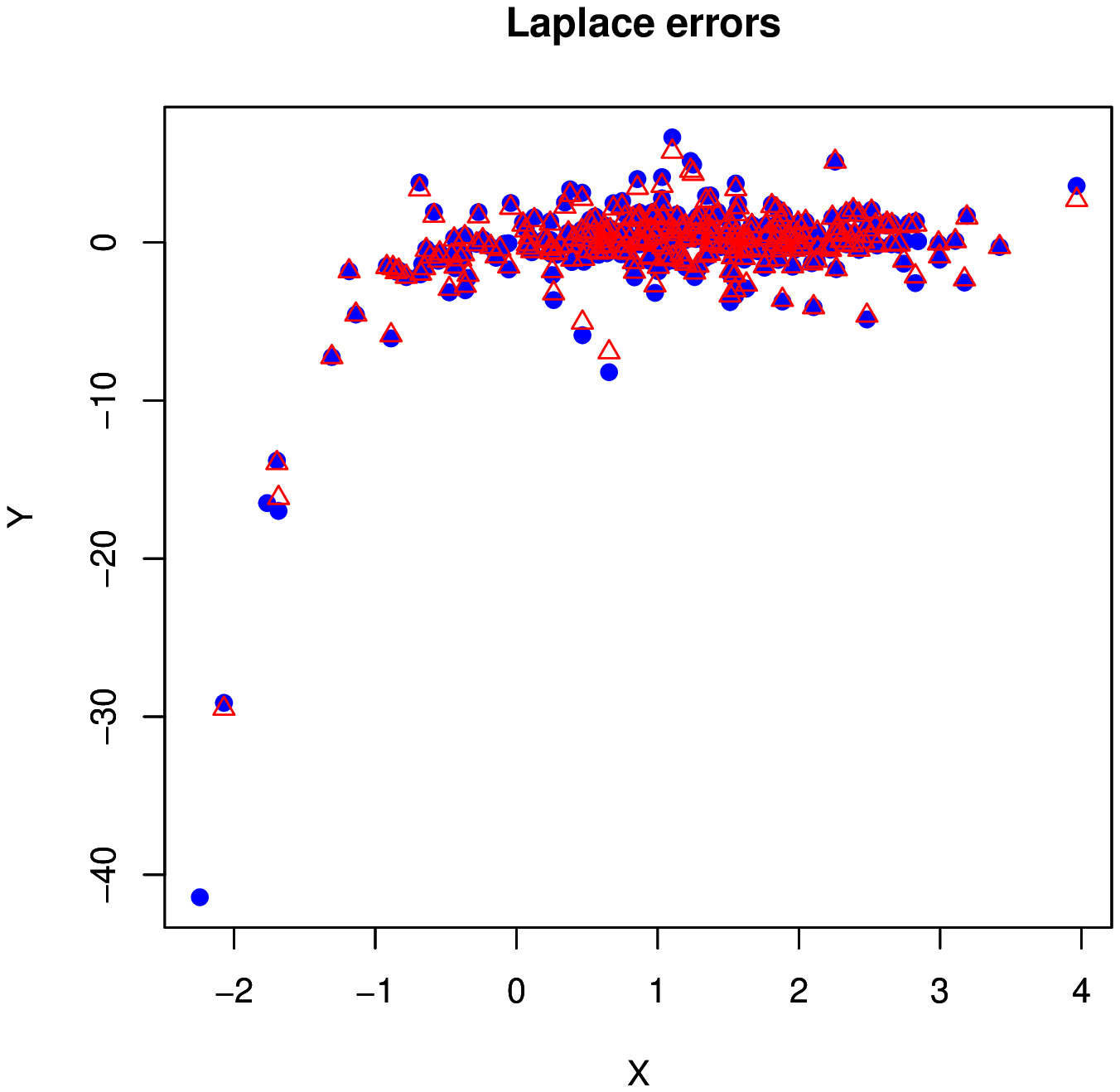}
\caption{\it Response variable $Y$ function regressor $X$: ``solide circle'' for reconstituted values $Y_{n,i}$, ``triangle'' for true complete values $Y_i$. Laplace error $\varepsilon \sim {\cal L}(0,1)$, n=300.}
\label{Figure 2}
\end{minipage}
\end{figure}

\begin{figure}[h!]
\begin{minipage}[b]{0.53\linewidth}
\includegraphics[scale=0.48]{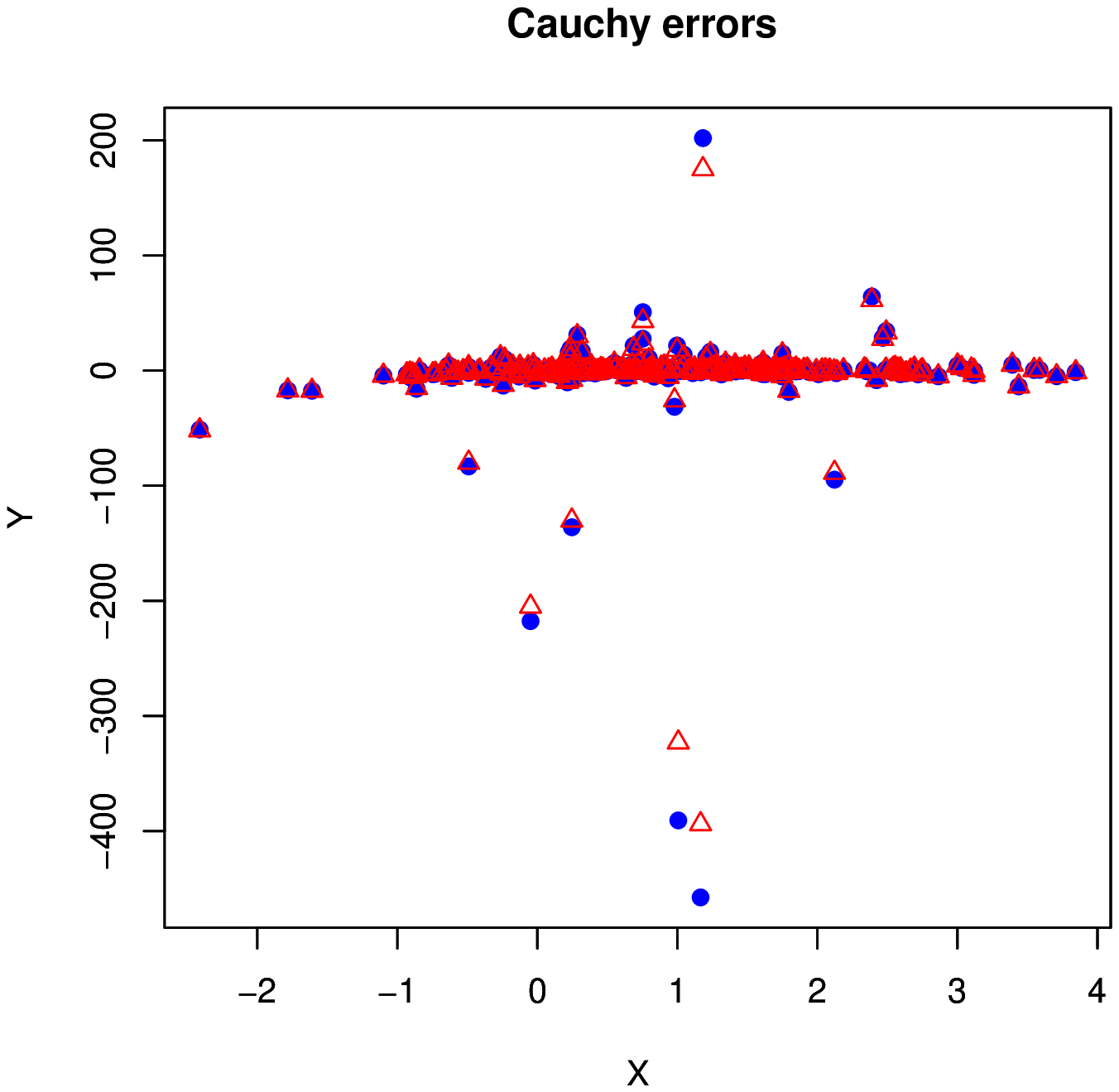} 
\caption{\it Response variable $Y$ function regressor $X$: ``solide circle'' for reconstituted values $Y_{n,i}$, ``triangle'' for true complete values $Y_i$. Cauchy error $\varepsilon \sim {\cal C}(0,1)$, n=300.}
\label{Figure 3}
\end{minipage}
\begin{minipage}[b]{0.53\linewidth}
\includegraphics[scale=0.48]{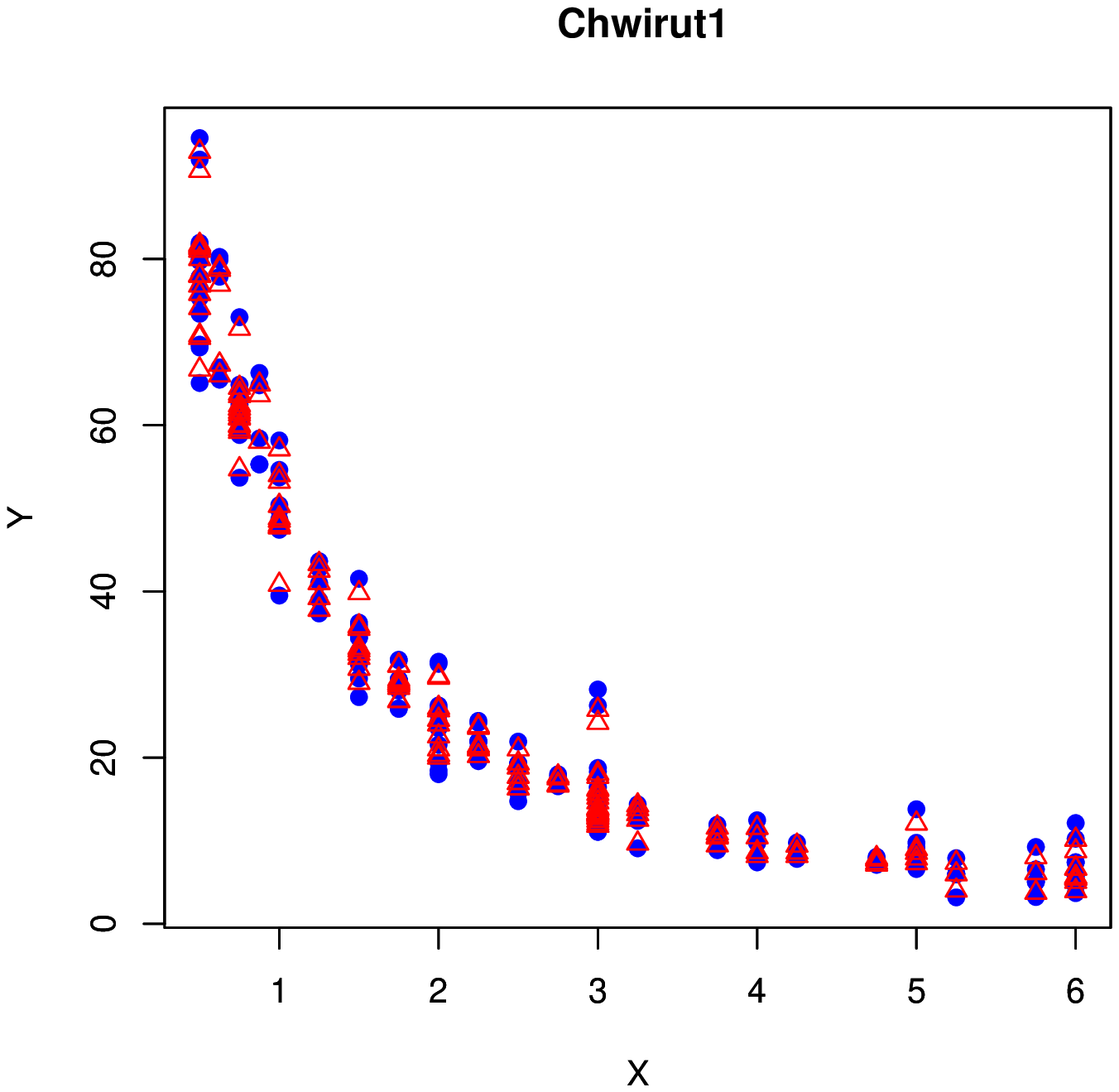}
\caption{\it Variable $Y$=ultrasonic response function regressor $X$=metal distance: ``solide circle'' for reconstituted values $Y_{n,i}$, ``triangle'' for true complete values $Y_i$.}
\label{Figure 4}
\end{minipage}
\end{figure}

Let us consider following nonlinear function corresponding a two-compartment model:
\begin{equation*}
f(x;\eb)=\frac{\beta_1}{\beta_1-\beta_2}(e^{-\beta_2 x}-e^{-\beta_1 x}), \qquad \eb=(\beta_1,\beta_2),
\end{equation*}
with the true values: $\beta_1^0=1$ and $\beta_2^0=1.5$. \\
In Figures \ref{Figure 1}, \ref{Figure 2}, \ref{Figure 3} a simulation of this model is plotted  for $n=300$, $\pi(x)=0.8$, $\sigma=1$ for  errors $\varepsilon \sim {\cal N, L, C}$ respectively. We represented  with ``solid circle'' the reconstituted values $Y_{n,i}$ and with ``triangle'' the true complete values $Y_i$. \\
In Table \ref{tableau1} we get the coverage probability  $CP_{LS}$, $CP_{LAD}$, ${\hat {CP}_{LS}}$, $NCP_{LS}$, $NCP_{LAD}$ in the case $\pi(X)=0.8+0.2|x-1|$ if $|x-1| \leq 1$ and 0.95 elsewhere when $n=300$, $n=100$. In Table \ref{tableau2} we get the same five coverage probabilities for the case $\pi(X)=0.8$. These two tables show that:
\begin{enumerate}
 \item $CP_{LS}$ is better than $CP_{LAD}$ excepting for the case  $\varepsilon \sim {\cal C}(0,2)$;
\item coverage probabilities on the reconstituted responses $\hat CP_{LS}$ is  bigger than 0.95 and has values very closed to values   of  $CP_{LS}$ obtained on complete data; 
\item $NCP_{LS}$, $NCP_{LAD}$ are worse than coverage rate on the complete or reconstituted data, particularly for  $n=100$ or Cauchy errors. For Cauchy distribution  we cannot calculate $NCP_{LS}$ since the assumptions of  Theorem \ref{theorem2}(i), more precisely  (H1), are not satisfied ;
\item Since case a) has tendency to eliminate the outliers of $\varepsilon$, the rates of normal coverage probability  are bigger for a) than for b).
\end{enumerate}
Tables \ref{tableau3} and  \ref{tableau3bis} contain empirical mean, standard-deviation of parameter estimations $\hat \eb_{n;LS}$ and $\hat \eb_{n;LAD}$ for the two cases of response probability. We deduce from these two tables:
\begin{enumerate}
 \item there is no difference between the parameter  estimations obtained for both cases of probability a) and b), for the same $n$  and the same distribution of $\varepsilon$;
\item already known thing: for distributions with outliers (Laplace or Cauchy here) the LAD estimators have a standard-deviation smaller than LS estimators.
\end{enumerate}
Table \ref{tableau4}: whether it is for the probability a) or b), the medians of  LAD estimations  are more close to the true values than the  medians of LS estimations, if $\varepsilon$  have  outliers. 
The  precision of the estimators increases with $n$ and in a less measure with the  probability $\pi(x)$.  
 
\subsection{Linear model}
Consider now the simple linear model: $Y_i=X_i\beta+\varepsilon_i$, with the  true value of the parameter $\beta^0=10$.  
In Tables \ref{tableau1linear} and \ref{tableau2linear}  we get the coverage probabilities for the two case of $\pi(X)$ and for $n=100,20$. From these two tables  we make the following remarks:
\begin{enumerate}
 \item in general $CP_{LAD}$ performs better than  $CP_{LS}$ when $\varepsilon \sim {\cal L}$ or ${\cal C}$;
\item $\hat CP_{LS}$ is bigger than 0.95 and has very close values to $CP_{LS}$ values;
\item similar of (3) for nonlinear case, except that $NCP_{LAD}< NCP_{LS}$;
\item with regard to both cases of probability $\pi(X)$:  there is no difference between the three coverage probabilities by estimated EL methods,  contrary to the nonlinear case nor between the natural coverage probabilities.  
\end{enumerate}
Tables \ref{tableau3linear} and \ref{tableau4linear}:  the standard-deviation of LS  estimators  is much bigger than for LAD's when $\varepsilon \sim {\cal C}$. That comes from the  outliers presence in  nonlinear case, the  algorithm to find the LS estimator can not converge, thus the estimations are not available.  
\subsection{Conclusion}
The coverage probabilities of CR given by normal method are lower than the nominal level $1-\alpha$ especially for small sample sizes. The simulation results show that in terms of coverage probability the empirical likelihood methods outperform the normal approximation method in particular when error distribution have outliers or its has a bigger standard-deviation. \\
It is interesting to note that, if the distribution of the errors presents outliers the approximated EL for LAD method gives generally  best results than that for LS method that concerns EL coverage probability or parameter estimators efficiency. Then, if $\varepsilon \sim {\cal N}$, in order to test hypothesis  $H_0: \eb=\ebo$, it is recommended to take the test statistic  $ l^*_n(\eb)$ given by (\ref{e19}) for complete data or $\hat l^*_n(\eb)$, given by (\ref{e20}) for improved data using LS method. Let us emphasize that $\hat l^*_n(\eb)$ gives very close results to those of $l^*_n(\eb)$. On the other hand, if $\varepsilon$ presents outliers,  to test $H_0: \eb=\ebo$ it is recommended $ l^*_n(\eb)$ given by (\ref{e19}) at $\eg_i=\eg_{i,LAD}$.
\section{An application}
In R language consider {\it Chwirut1} data of {\it NISTnls} package to model the ultrasonic response value $Y$ function to metal distance value $X$ using nonlinear function:
\begin{equation*}
f(X,\eb)=\frac{\exp(-b_1X)}{b_2+b_3X}, \qquad \eb=(b_1,b_2, b_3).
\end{equation*}
The realizations of $(Y,X)$ are known for $n=214$ observations. In Figure \ref{Figure 4} we represented variable $Y$=ultrasonic response function of regressor $X$=metal distance: ``solid circle'' for reconstituted values $Y_{n,i}$, ``triangle'' for true complete values $Y_i$. The LS parameter estimations  on all 214 observations are: $\hat b_{1}=0.19$, $\hat b_{2}=0.006$, $\hat b_{3}=0.10$. By 10000 Monte Carlo samples, we eliminate $20\%$, $50\%$, $80\%$ values of $Y$. Consider that the true value $\ebo$ of $\eb$ is  the one obtained on 214 observations.  In Table \ref{tableau6} we have  the acceptance rate of hypothesis $H_0: \eb=\ebo$ with respect to  missing probability $1-\pi(X)$ and with respect to the estimation method (LS or LAD) on complete data using statistic $ l^*_n(\eb)$ given by (\ref{e19}), on reconstituted data using statistic $\hat l^*_n(\eb)$ given by (\ref{e20}). We observe that,  even by eliminating $80\%$ observations, the obtained estimations are very close to $\ebo$. In Table \ref{tableau7}   we find the empirical means, standard-deviations of differences $Y_i - \hat Y_{i,0}$, $Y_i - \hat Y_{i}$, where: $Y_i$ are the true values of $Y$, $\hat Y_{i,0}$   is the forecast for $Y_i$ using all 214 observations by LS method, $\hat Y_{i,0}=f(X_i,\hat \eb_{LS}) $. For $i=1, \cdots, n$, $\hat Y_{i}=Y_{ni}$   is the reconstituted  value of $Y_i$ by relation (\ref{yy}). We deduce that the variable response is better reconstituted by (\ref{yy}) than by LS method on all observations. 

\section{Appendix: proofs}
We give proofs for the results in Sections 3 and 4.\\
In the following, we denote by $C$ a generic positive finite constant not depending on $n$ which may take different values in different formulae or even in different parts of the same formula.\\

\noindent {\bf Proof of Lemma \ref{lemma2}}
\textit{(i)} We apply the central limit theorem. \\
\textit{(ii)} By the weak law of large numbers.\\
\textit{(iii)} We combine Lemma 3 of \cite{Owen:90} and results for nonlinear regression without missing data (see e.g. \cite{Seber:Wild:03}).
\hspace*{\fill}$\blacksquare$ \\

{\bf Proof of Theorem \ref{theorem1}}
We take the Taylor's expansion of $l_n(\ebo)$ given by (\ref{e7}): $
l_n(\ebo)=2 \el^t \sum^n_{i=1} \eg_i(\ebo)- \el^t \sum^n_{i=1} \eg_i(\ebo) \eg_i^t(\ebo) \el +  \el^t O_{\eP}(\sum^n_{i=1} \eg_i(\ebo) (\el^t \eg_i(\ebo) )^2)$. 
By a similar approach of \cite{Owen:90}, using also Lemma \ref{lemma2},  we obtain: $\| \el \| =O_{\eP}(n^{-1/2})$. 
Using Lemma \ref{lemma2}(iii) we obtain:
\begin{equation}
\label{e11}
 n^{-1} \el^t\sum^n_{i=1} \eg_i(\ebo) ( \el^t \eg_i(\ebo) )^2 =o_{\eP} \pth{n^{-1}  \sum^n_{i=1} ( \el^t \eg_i(\ebo))^2}.
\end{equation}
Then, returning at $l_n(\ebo)$, we have:
\begin{equation}
 \label{e10}
l_n(\ebo)=2 \sum^n_{i=1} \cro{ \el^t \eg_i(\ebo)- \frac{1}{2}  \el^t \eg_i(\ebo) \eg_i^t(\ebo) \el}(1+o_{\eP}(1)).
\end{equation}
 Relation (\ref{e8}) can be written also:
$$
\textbf{0}=\frac{1}{n} \sum^n_{i=1} \eg_i(\ebo) -\frac{1}{n} \sum^n_{i=1}  \eg_i(\ebo)\eg_i^t(\ebo) \el +\frac{1}{n} \sum^n_{i=1} \eg_i(\ebo)
( \el^t \eg_i(\ebo))^2 (1+\el^t \eg_i(\ebo))^{-1}.
$$
On the other hand, using (\ref{e11}), we obtain:
\begin{equation}
 \label{e12}
\textbf{0}=\cro{\frac{1}{n} \sum^n_{i=1} \eg_i(\ebo) - \frac{1}{n} \sum^n_{i=1} \eg_i(\ebo)\eg_i^t(\ebo) \el } (1+o_{\eP}(1)).
\end{equation}
Thus
\begin{equation}
 \label{e13}
\el=\cro{\frac{1}{n} \sum^n_{i=1} \eg_i(\ebo)\eg_i^t(\ebo)}^{-1} \cro{\frac 1n \sum^n_{i=1} \eg_i(\ebo)}(1+o_{\eP}(1)).
\end{equation}
Relation (\ref{e12}) implies also: $\el^t \sum^n_{i=1} \eg_i(\ebo)=\cro{ \el^t \sum^n_{i=1} \eg_i(\ebo) \eg_i^t(\ebo) \el }(1+o_{\eP}(1))$. 
Then, expression (\ref{e10}) of $l_n(\ebo)$ becomes, using relation (\ref{e13}):
\begin{equation*}
l_n(\ebo)=\cro{\el^t \sum^n_{i=1}\eg_i(\ebo)}(1+o_{\eP}(1))= \textbf{B}_n^{-1} \pth{\frac{1}{\sqrt n} \sum^n_{i=1}  \eg_i^t(\ebo) } \pth{\frac{1}{\sqrt n} \sum^n_{i=1}  \eg_i(\ebo) }(1+o_{\eP}(1)).
\end{equation*}
By Lemma \ref{lemma2}(i) and (ii) we have: $l_n(\ebo) \overset{{\cal L}} {\underset{n \rightarrow \infty}{\longrightarrow}} \chi^2(d)$.
\hspace*{\fill}$\blacksquare$ \\

{\bf Proof of Theorem \ref{theorem2}}.(i)
By definition, the LS estimator is obtained as the solution of the system: $\textbf{0}= \sum^n_{i=1} \eg_i(\hat \eb_{n,LS})$. Using Lemma \ref{lemma2}(i) we have: $
n^{-1/2} \sum^n_{i=1} \eg_i(\ebo)\overset{{\cal L}} {\underset{n \rightarrow \infty}{\longrightarrow}}{\cal N}_d(\textbf{0}, \textbf{B})$, then $
n^{-1/2} \sum^n_{i=1} [\eg_i(\hat \eb_{n,LS})-\eg_i(\ebo)]\overset{{\cal L}} {\underset{n \rightarrow \infty}{\longrightarrow}}{\cal N}(0, \textbf{B})$. 
This last relation implies:
\begin{equation}
 \label{e14}
n^{-1} \sum^n_{i=1} [\eg_i(\hat \eb_{n,LS})-\eg_i(\ebo)]=O_{\eP}(n^{-1/2}).
\end{equation}
On the other hand, using Taylor expansion:\\
$
\eg_i(\hat \eb_{n,LS})-\eg_i(\ebo)=\delta_i \ef(\eX_i;\hat \eb_{n,LS})[Y_i-f(\eX_i;\hat \eb_{n,LS})]-\delta_i \ef(\eX_i;\ebo)[Y_i-f(\eX_i;\ebo)]
$\\
$
=\delta_i \varepsilon_i \cro{\ef(\eX_i;\hat \eb_{n,LS})-\ef(\eX_i;\ebo)}+\delta_i \cro{\ef(\eX_i;\hat \eb_{n,LS})-\ef(\eX_i;\ebo)} \cro{f(\eX_i,\ebo)-f(\eX_i;\hat \eb_{n;LS})}$. \\
\hspace*{1cm} $+\delta_i \ef(\eX_i;\ebo)\cro{f(\eX_i,\ebo)-f(\eX_i;\hat \eb_{n;LS})}$\\
Since $\hat \eb_{n,LS}\overset{{a.s.}} {\underset{n \rightarrow \infty}{\longrightarrow}} \ebo$,  we obtain: $
 \eg_i(\hat \eb_{n,LS})-\eg_i(\ebo)=\delta_i \varepsilon_i \eff(\eX_i;\eb_{1,n})(\hat \eb_{n,LS}-\ebo)$
 $
+(\hat \eb_{n,LS}-\ebo)\delta_i\ef^t(\eX_i;{\eb_{2,n}})\eff(\eX_i;\eb_{1,n})(\hat \eb_{n,LS}-\ebo)
+(\hat \eb_{n,LS}-\ebo)\delta_i\ef^t(\eX_i;{\eb_{2,n}})  \ef(\eX_i;\ebo)
 $, 
with $\eb_{1,n}=\ebo+u_1(\hat \eb_{n,LS}-\ebo)$,  $\eb_{2,n}=\ebo+u_2(\hat \eb_{n,LS}-\ebo)$, $u_1,u_2 \in [0,1]$.\\
Under assumptions (H1) and (H4): \\
$n^{-1}\sum^n_{i=1}\cro{\eg_i(\hat \eb_{n,LS})-\eg_i(\ebo)}=\cro{o_{\eP}(\hat \eb_{n,LS}-\ebo)+(\hat \eb_{n,LS}-\ebo)A}(1+o_{\eP}(1))$. Taking into account relation (\ref{e14}), we have:
\begin{equation}
 \label{e15}
\hat \eb_{n,LS}-\ebo=O_{\eP}(n^{-1/2}).
\end{equation}
Combining these last two results, we obtain: 
$$
{\sqrt n} (\hat \eb_{n,LS}- \ebo) A (1+o_{\eP}(1)) =n^{-1/2} \sum^n_{i=1} \cro{\eg_i(\hat \eb_{n,LS}-\eg_i(\ebo)) }\overset{{\cal L}} {\underset{n \rightarrow \infty}{\longrightarrow}}{\cal N}_d(\textbf{0}, \textbf{B})
$$ 
and claim follows.\\
(ii) We apply \cite{Weiss:91}.
\hspace*{\fill}$\blacksquare$ \\

\noindent {\bf Proof of Lemma \ref{lemma3}}
\textit{(i)} 
Let us consider following decomposition: \\
$\frac{1}{\sqrt n} \sum^n_{i=1} (Y_{n,i}-\theta^0)\equiv S_1+S_2+S_3$, with:\\
$S_1=\frac{1}{\sqrt n} \sum^n_{i=1} \cro{\frac{\delta_i\varepsilon_i}{\pi(\eX_i)}+f(\eX_i;\ebo)-\theta^0}$,\\
$S_2=\frac{1}{\sqrt n} \sum^n_{i=1} \delta_i\varepsilon_i \cro{\frac{1}{\hat \pi(\eX_i)}- \frac{1}{\pi(\eX_i)}}$,\\
$S_3=\frac{1}{\sqrt n} \sum^n_{i=1} \cro{f(\eX_i;\hat \eb_{n,LS})-f(\eX_i;\ebo)} \pth{1-\frac{\delta_i}{\hat \pi(\eX_i)}}$.\\
\underline{For  $S_1$}: $Var[\frac{\delta_i \varepsilon_i}{\pi(\eX_i)}+f(\eX_i;\ebo)-\theta^0]=Var[\frac{\delta_i \varepsilon_i}{\pi(\eX_i)}]+Var[f(\eX_i;\ebo)]+2Cov(f(\eX_i;\ebo),\frac{\delta_i \varepsilon_i}{\pi(\eX_i)})$. By (H1) we have $Cov(f(\eX_i;\ebo),\frac{\delta_i \varepsilon_i}{\pi(\eX_i)})=0$. On the other hand: $
Var\left[\frac{\delta_i \varepsilon_i}{\pi(\eX_i)}\right]=\eE\left[\frac{\delta_i^2 \varepsilon_i^2}{\pi^2(\eX_i)}\right]=\eE\left[\frac{\delta_i \varepsilon_i^2}{\pi^2(\eX_i)}\right]=\eE\left[\frac{\varepsilon^2(\eX)}{\pi(\eX)}\right]$. 
Since $(\eX_i,\varepsilon_i)$ are  independent for different $i$, then the random variables $\frac{\delta_i \varepsilon_i}{\pi(\eX_i)}-f(\eX_i;\ebo)-\theta^0$ are also independent. Then, we can apply the central limit theorem: $S_1 \overset{{\cal L}} {\underset{n \rightarrow \infty}{\longrightarrow}}{\cal N}(0, W)$.\\
\underline{For  $S_3$}. 
We make a  limited development for $S_3$   until order 2 around $\ebo$ and taking into account relation (\ref{e15}), hypothesis $\pi(\ex)>0$ and (H4): \\
$S_3=(\hat \eb_{n,LS}-\ebo)^t\frac{1}{\sqrt n}\sum^n_{i=1} \pth{1-\frac{\delta_i}{\pi(\eX_i)}} \ef(\eX_i;\ebo)(1+o_{\eP}(1))=O_{\eP}(n^{-1/2})o_{\eP}(n^{1/2})=o_{\eP}(1)$. 
By Lemma 3 of \cite{Xue:09} we have $S_2=o_{\eP}(1)$, then the claim (i) is proved. The proof of (ii) and (iii) is similar.
\hspace*{\fill}$\blacksquare$ \\

\noindent {\bf Proof of Lemma \ref{lemma4}}
\textit{(i)} Function $g_{n,i}(\eb)$ can be written:
\begin{equation}
 \label{e16}
\eg_{n,i}(\eb)=\frac{\eg_{i,LS}(\eb)}{\hat \pi(\eX_i)} +\pth{1-\frac{\delta_i}{\hat \pi(\eX_i)}} [f(\eX_i;\hat \eb_{n,LS})-f(\eX_i;\eb)] \ef(\eX_i;\eb).
\end{equation}

For the first term of the right-hand side  of (\ref{e16}) we have:
\begin{equation}
 \label{e17}
 n^{-1/2} \sum^n_{i=1} \frac{\eg_{i,LS}(\ebo) }{\hat \pi(\eX_i)}  \overset{{\cal L}} {\underset{n \rightarrow \infty}{\longrightarrow}}{\cal N}_d \pth{\textbf{0}, \eE \cro{\pi(\eX)^{-1} \varepsilon^2(\eX) \ef(\eX;\ebo) \ef^t(\eX;\ebo)}}.
\end{equation}
On the other hand, using a  limited development, relation (\ref{e15}) and assumption (H4): $
n^{-1/2} \sum^n_{i=1}\pth{1-\frac{\delta_i}{\hat \pi(\eX_i)}}[f(\eX_i;\hat \eb_{n,LS})-f(\eX_i;\ebo)] \ef(\eX_i;\ebo)$\\ $
=\frac{\hat \eb_{n,LS}- \ebo}{\sqrt n} \sum^n_{i=1}\pth{1-\frac{\delta_i}{\hat \pi(\eX_i)}} \ef^t(\eX_i;\eb_{1,n})\ef(\eX_i;\ebo)$, 
with  $\eb_{1,n}$ between $\ebo$ and $\hat \eb_{n,LS}$. Let us consider following random variable:\\
 $S_n(\eb)=n^{-1} \sum^n_{i=1}\pth{1-\frac{\delta_i}{\hat \pi(\eX_i)}} \ef^t(\eX_i; \eb)\ef(\eX_i;\ebo)$, for $\eb$ in a $n^{-1/2}$-neighborhood of $\ebo$. If we prove:
\begin{equation}
 \label{*0} S_n(\eb)=o_{\eP}(1),
\end{equation}
with $o_{\eP}(1)$ uniformly in $\eb$, tacking into account also (\ref{e15}), we obtain (i). \\
Let us remind at first some results of the paper \cite{Xue:09}, on the estimators $\hat \pi(\eX_i)$ of $\pi(\eX_i)$. Under assumptions (H3), (H6) and (H7) we have uniformly over $1 \leq i \leq n$:
\begin{equation}
 \label{elemma1}
\eE[\hat \pi(\eX_i)-\pi(\eX_i)]^2=O((nh_n^d)^{-1}M_n^d)+O(h_n^{2\max(2,d-1)})+o(n^{-1/2}).
\end{equation}
In the same paper, following two random variables are considered: $W_{nj}(\ex)=K_h(\eX_k-\ex)/ \max\{1, \sum^n_{i=1}K_h(\eX_i-\ex) \}$, for $\ex \in \Upsilon$, $j=1, \cdots, n$ and $C_n(\eX_i)=\max \{1, C_1 \sum_{j \neq i} \e1_{\| \eX_j-\eX_i \| \leq  \rho h_n}   \}$ with $C_1$, $\rho$ positive constants defined in assumption (H6).    For this last random variable we have $\eE\cro{C_n^{-1}(\eX_i)}=o_{\eP}(n^{-1/2})$.\\
We now turn to study $S_n(\eb)$, which can be written, after a limited development:
\begin{equation}
 \label{*1}
S_n(\eb)=\frac{1}{n} \sum^n_{i=1} \pth{1-\frac{\delta_i}{\hat \pi(\eX_i)}} \| \ef(\eX_i;\ebo) \| ^2 +\frac{\eb -\ebo}{n} \sum^n_{i=1} \pth{1-\frac{\delta_i}{\hat \pi(\eX_i)}} \eff(\eX_i; \tilde \eb) \ef(\eX_i; \ebo)
\end{equation}
with $\tilde \eb$ between $\ebo$ and $\eb$. We denote decomposition (\ref{*1}): $S_n(\eb)\equiv S_1+S_2$. Using Cauchy-Schwarz inequality, relation (\ref{e15}) and assumption (H4), it is easily shown that $S_2=o_{\eP}(1)$ uniformly in $\eb$. We study now $S_1=\frac{1}{n} \sum^n_{i=1} \pth{1- \frac{\delta_i}{\pi(\eX_i)}} \| \ef(\eX_i;\ebo)\|^2+\frac{1}{n} \sum^n_{i=1} \delta_i \pth{\frac{1}{\pi(\eX_i)}-\frac{1}{\hat \pi(\eX_i)}} \|\eff(\eX_i;\ebo) \|^2 \equiv S_{11}+S_{12}$.  Second term $S_{12}$ can be written: $ S_{12}=n^{-1} \sum^n_{i=1} \delta_i \frac{\hat \pi(\eX_i)-\pi(\eX_i)}{\pi^2(\eX_i)} \|\ef(\eX_i;\ebo) \|^2 - n^{-1}  \sum^n_{i=1} \frac{\pth{\hat \pi(\eX_i)-\pi(\eX_i)}^2}{\pi^2(\eX_i) \hat \pi(\eX_i) }\|\ef(\eX_i;\ebo) \|^2 $. The last term of $S_{12}$ is $o_{\eP}(1)$ by (\ref{elemma1}). For the first term of $S_{12}$ we have the decomposition:\\ $n^{-1} \sum^n_{i=1} \delta_i \frac{\hat \pi(\eX_i)-\pi(\eX_i)}{\pi^2(\eX_i)} \|\ef(\eX_i;\ebo) \|^2 \equiv T_1 +T_2 +T_3$, with \\$T_1= n^{-1} \sum^n_{i=1} \frac{\delta_i}{\pi^2(\eX_i)} \|\ef(\eX_i;\ebo) \|^2 \sum^n_{j=1} W_{nj}(\eX_i)[\pi(\eX_i)-\pi(\eX_j)] $,\\ $T_2=n^{-1} \sum^n_{i=1} \frac{\delta_i}{\pi^2(\eX_i)} \|\ef(\eX_i;\ebo) \|^2 \sum^n_{j=1} W_{nj}(\eX_i)[\pi(\eX_j)- \delta_j]$,\\
  $T_3=n^{-1} \sum^n_{i=1} \frac{\delta_i}{\pi(\eX_i)} \|\ef(\eX_i;\ebo) \|^2 \cro{1- \sum^n_{j=1} W_{nj}(\eX_i)}$. \\
By Cauchy-Schwarz inequality, assumptions (H4), (H7) we have:\\
 $\eE[T^2_1] \leq \frac{C}{n^2} \sum^n_{i=1} \eE \cro{  \eE\cro{ \sum^n_{j=1} W^2_{nj}(\eX_i)[\pi(\eX_i)-\pi(\eX_j)]^2 | \eX_i  } }$ \\  $\leq \frac{C}{n^2} \sum^n_{i=1} \eE \cro{  \eE\cro{ \sum^n_{j=1} W^2_{nj}(\eX_i)[\eX_i-\eX_j]^2 | \eX_i  } } $ $\leq Ch^2_n n^{-1} \rightarrow 0 $ for $n \rightarrow \infty$. Then $T_1=o_{\eP}(1)$.\\ For $T_2$: $\eE[T_2] \leq \frac{C}{n^2} \sum^n_{i=1} \eE \cro{\sum^n_{j=1} W^2_{nj}(\eX_i) } \leq \frac{C}{n^2} \sum^n_{i=1} \eE \cro{\frac{1}{C_n(\eX_i)}}\rightarrow 0 $ for $n \rightarrow \infty$, then $T_2=o_{\eP}(1)$. In a similar way:\\
 $\eE[T_3] \leq \frac{C}{n^2} \sum^n_{i=1} \cro{\eE \pth{\frac{\delta_i}{\pi^2(\eX_i)} \| \ef(\eX_i;\ebo) \|^2 |\eX_i } \pth{1-\sum^n_{j=1} W_{nj}(\eX_i) }} \rightarrow 0$. Thus $T_1+T_2+T_3=o_{\eP}(1)$, what implies $S_{12}=o_{\eP}(1)$. \\
Finally, for $S_{11}$: $\eE[S_{11}]=0$ and with the same arguments as for $T_2$: $\eE[S^2_{11}]\rightarrow 0$ for $n\rightarrow n$, then $S_{11}=o_{\eP}(1)$.  With all this, relation (\ref{*0}) is proved. \\
the proof of (ii) and (iii) is similar.
\hspace*{\fill}$\blacksquare$ \\

~\newpage

 \newpage
\begin{table}
\caption{Coverage probabilities for nonlinear model, $\pi(X)=0.8+0.2|x-1|$, if $|x-1| \leq 1$ and 0.95 elsewhere, $n=300,100$, error Gaussian, Laplace, Cauchy. }
\scriptsize{
\begin{tabular}{|c|ccccc|ccccc|} \hline
   &  &$n=$ & $300$&  & &  &$n=$ & $100$& &   \\
 $\varepsilon$    &  $CP_{LS}$ &  $CP_{LAD}$ & ${\hat {CP}_{LS}}$ &  $NCP_{LS}$ &  $NCP_{LAD}$ &$CP_{LS}$ &  $CP_{LAD}$ & ${\hat {CP}_{LS}}$&  $NCP_{LS}$ &  $NCP_{LAD}$ \\ \hline
${\cal N}(0,1)$ &   0.977 &  0.825 & 0.959 &  0.981 & 0.884 & 0.975 & 0.747 & 0.956 &  0.63 & 0.507  \\
${\cal L}(0,1)$ &   0.971 & 0.881 & 0.967 & 0.991 & 0.829 & 0.978 & 0.836 & 0.95 & 0.604 & 0.453 \\ 
${\cal C}(0,1)$ &  0.983 & 0.991 &  0.984 &    & 0.293 & 0.987  & 0.963 & 0.987 &   & 0.586  \\ 
\hline
${\cal N}(0,2)$ & 0.964 & 0.964 &  0.945 &  0.774 & 0.423 & 0.976 & 0.957 & 0.917 &  0.86 & 0.686 \\
${\cal L}(0,2)$ &  0.977 & 0.979 &  0.974 & 0.455 & 0.394 & 0.978 & 0.959 & 0.978 & 0.925 & 0.439 \\ 
${\cal C}(0,2)$ & 0.986  & 0.999 & 0.988 &     & 0.615 & 0.987 & 0.994 & 0.987 &   & 0.541  \\ \hline
\end{tabular} 
}
\label{tableau1}
\end{table}

\begin{table}
\caption{Coverage probabilities for nonlinear model, $\pi(X)=0.8$, $n=300,100$, error Gaussian, Laplace, Cauchy. }
\scriptsize{
\begin{tabular}{|c|ccccc|ccccc|} \hline
   &  &$n=$ & $300$&  & &  &$n=$ & $100$& &   \\
 $\varepsilon$    &  $CP_{LS}$ &  $CP_{LAD}$ & ${\hat {CP}_{LS}}$ &   $NCP_{LS}$ &  $NCP_{LAD}$ &$CP_{LS}$ &  $CP_{LAD}$ & ${\hat {CP}_{LS}}$&   $NCP_{LS}$ &  $NCP_{LAD}$ \\ \hline
${\cal N}(0,1)$ &  0.970 & 0.807 & 0.960 & 0.955 & 0.923& 0.972 & 0.780 & 0.931  & 0.508 & 0.393 \\
${\cal L}(0,1)$ & 0.972 & 0.872 & 0.975 &0.986 & 0.891 & 0.976  & 0.823 & 0.975 &  0.227  & 0.392 \\ 
${\cal C}(0,1)$ &  0.985 & 0.985 &  0.987 &   & 0.753 & 0.987 & 0.958 & 0.987 &   & 0.772  \\ \hline
${\cal N}(0,2)$& 0.979 & 0.968 & 0.949 &0.560 & 0.298 & 0.979   & 0.958  & 0.886&0.159  & 0.254   \\
${\cal L}(0,2)$ &  0.979 &0.977 &  0.969 &  0.668 & 0.288 & 0.979 & 0.965 & 0.973 &  0.353 & 0.234 \\ 
${\cal C}(0,2)$ & 0.986  & 0.999 & 0.985  &   & 0.636 & 0.991 & 0.992 & 0.990  & & 0.699  \\ \hline
\end{tabular} 
}
\label{tableau2}
\end{table}

\begin{table}
\caption{Mean, standard deviation of parameter estimations for nonlinear model, $\pi(X)=0.8+0.2|x-1|$ if $|x-1| \leq 1$ and 0.95 elsewhere, error Gaussian, Laplace, Cauchy. True values $\beta^0_1=1$, $\beta^0_2=1.5$.}
\scriptsize{
\begin{tabular}{|c||cc|cc||cc|cc||} \hline
 &  & $\beta_1$ & & & & $\beta_2$ & & \\
$n$ & mean & mean & sd & sd & mean & mean & sd & sd \\
error & LS & LAD & LS & LAD &  LS & LAD & LS & LAD \\ \hline
$n=300$ &  &  & & & &  & &  \\ \hline
 ${\cal N}(0,1)$ &  1& 1 &0.13 & 0.16 & 1.50 & 1.49 &0.26 & 0.31 \\
${\cal L}(0,1)$ & 1 & 1 & 0.17 & 0.14 & 1.48 & 1.50  & 0.36 & 0.28 \\ 
 ${\cal C}(0,1)$ &  0.99 & 0.99 &0.68 & 0.15 & 1.40 & 1.50  & 1.58 & 0.33 \\
${\cal N}(0,2)$ & 1  & 0.99 & 0.25 & 0.28 & 1.49 & 1.51 & 0.53 & 0.62 \\
${\cal L}(0,2)$ & 0.99 & 0.98 & 0.31 & 0.26 & 1.51 & 1.52  & 0.74 & 0.61 \\
${\cal C}(0,2)$ & 1.01 & 0.95 &0.82 & 0.36 & 1.43 & 1.61  & 2.06 & 1.02 \\ \hline
$n=100$ &  &  & & & &  & &  \\ \hline
 ${\cal N}(0,1)$ &  0.97& 0.96 &0.25 & 0.29 & 1.58 & 1.59 &0.70 & 0.77 \\
${\cal L}(0,1)$ & 0.96 & 0.95 & 0.33 & 0.29 & 1.60 & 1.61  & 0.97 & 0.86 \\ 
${\cal C}(0,1)$ &  1.01 & 0.94 &0.79 & 0.83 & 1.54 & 1.66  & 2.32 & 2.09 \\
 ${\cal N}(0,2)$ & 0.96  & 0.94 & 0.42 & 0.54  & 1.66  & 1.70  & 1.40 & 1.71  \\
${\cal L}(0,2)$ & 0.93  & 0.91 & 0.48  & 0.65 & 1.74  & 1.81  & 1.65 & 1.93 \\ 
${\cal C}(0,2)$ & 1.02 & 0.87 &0.93 & 0.90 & 1.53 & 1.85  & 2.53 & 2.21 \\ \hline 
\end{tabular} 
}
\label{tableau3}
\end{table}

\begin{table}
\caption{Mean, standard deviation of parameter  estimations for nonlinear model, $\pi(X)=0.8$, error Gaussian, Laplace, Cauchy. True values $\beta^0_1=1$, $\beta^0_2=1.5$.}
\scriptsize{
\begin{tabular}{|c||cc|cc||cc|cc||} \hline
 &  & $\beta_1$ & & & & $\beta_2$ & & \\
$n$ & mean & mean & sd & sd & mean & mean & sd & sd \\
error & LS & LAD & LS & LAD &  LS & LAD & LS & LAD \\ \hline
$n=300$ &  &  & & & &  & &  \\ \hline
 ${\cal N}(0,1)$ &  1& 0.99 &0.14 & 0.18 & 1.50 & 1.50 &0.29 & 0.35 \\
${\cal L}(0,1)$ & 1 & 0.99 & 0.19 & 0.15 & 1.50 & 1.50  & 0.40 & 0.31 \\
${\cal C}(0,1)$ &  0.99 & 0.96 &0.70 & 0.25 & 1.51 & 1.58  & 1.87 & 1.69 \\
${\cal N}(0,2)$ &  1 & 0.98 & 0.26& 0.31 & 1.49 & 1.52 & 0.58 & 0.70 \\
${\cal L}(0,2)$ & 0.97 & 0.96 & 0.34 & 0.29 & 1.54 & 1.57  & 0.82 & 0.72 \\ 
${\cal C}(0,2)$ & 0.96 & 0.96 &0.83 & 0.37 & 1.57 & 1.55  & 2.13 & 0.95  \\ \hline 
$n=100$ &  &  & & & &  & &  \\ \hline
 ${\cal N}(0,1)$ &  0.96& 0.95 &0.28 & 0.35 & 1.62 & 1.66 &0.99 & 1.21 \\
${\cal L}(0,1)$ & 0.96 & 0.95 & 0.32 & 0.31 & 1.62 & 1.64  & 1.23 & 0.94 \\ 
${\cal C}(0,1)$ &  0.97 & 0.95 &0.81 & 0.99 & 1.59 & 1.73  & 2.32 & 1.98 \\
 ${\cal N}(0,2)$ & 0.94  & 0.94 & 0.44 & 0.52  & 1.70  & 1.77  & 1.51 & 1.85 \\
${\cal L}(0,2)$ & 0.95  & 0.93 & 0.51  & 0.51 & 1.66  & 1.75  & 1.66 & 1.65 \\ 
${\cal C}(0,2)$ &  1.09 & 0.89 &0.94 & 0.58 & 1.25 & 1.83  & 2.28 & 2.04 \\ \hline 
\end{tabular} 
}
\label{tableau3bis}
\end{table}

\begin{table}
\caption{Median of parameter estimations for nonlinear model, two cases for $\pi(X)$, error Gaussian, Laplace or Cauchy.  True values $\beta^0_1=1$, $\beta^0_2=1.5$.}
\scriptsize{
\begin{tabular}{|c||cc|cc||cc|cc||} \hline
&  & $\beta_1$ & & & & $\beta_2$ & & \\
$n$ & case & a) & case & b) & case & a) & case & b) \\
error & LS & LAD & LS & LAD &  LS & LAD & LS & LAD \\ \hline
$n=300$ &  &  & & & &  & &  \\ \hline
 ${\cal N}(0,2)$ &  1& 1 & 1 & 1 & 1.51 & 1.5 &1.49  & 1.5 \\
${\cal L}(0,2)$ &  0.99 & 1 & 0.98 & 0.99 & 1.52 & 1.5 & 1.53 & 1.5 \\ 
${\cal C}(0,2)$ &  1.17 & 1 & 0.96 & 1  & 1.56 & 1.5 & 1.48 & 1.5\\ \hline  
$n=100$ &  &  & & & &  & &  \\ \hline
${\cal N}(0,2)$ &  0.97 & 0.96 & 0.99 & 0.94 & 1.45 & 1.5 & 1.48 & 1.51 \\
${\cal L}(0,2)$ &  0.99 & 0.99 & 1 &  0.99 & 1.39 & 1. 49 & 1.39 & 1.50 \\ 
${\cal C}(0,2)$ &  1.18 & 1 & 1.29 &  1 & 1.81 & 1.5 & 1.62 & 1.5 \\ \hline 
\end{tabular} 
}
\label{tableau4}
\end{table}

\begin{table}
\caption{Coverage probabilities for linear model, $\pi(X)=0.8+0.2|x-1|$, if $|x-1| \leq 1$ and 0.95 elsewhere, $n=100,20$, error Gaussian, Laplace, Cauchy. }
\scriptsize{
\begin{tabular}{|c|ccccc|ccccc|} \hline
  &  &$n=$ & $100$&  & &  &$n=$ & $20$& &   \\
  $\varepsilon$    &  $CP_{LS}$ &  $CP_{LAD}$ & ${\hat {CP}_{LS}}$ &   $NCP_{LS}$ &  $NCP_{LAD}$ &$CP_{LS}$ &  $CP_{LAD}$ & ${\hat {CP}_{LS}}$&   $NCP_{LS}$ &  $NCP_{LAD}$ \\ \hline
${\cal N}(0,1)$ &  0.95 & 0.943  & 0.954 &  0.945 & 0.802 &  0.967 &  0.921 & 0.955  &  0.994 & 0.785  \\
${\cal L}(0,1)$ &  0.960 & 0.991 &  0.960 &  0.962 &   0.784 & 0.965 & 0.959 & 0.965 &  0.864 & 0.757  \\
${\cal C}(0,1)$ &  0.985 & 1 &  0.986 &  & 0.745 &  0.984 & 0.999 & 0.979  &  & 0.737  \\ \hline
${\cal N}(0,2)$ & 0.956  & 1   & 0.956  & 0.983 & 0.631 & 0.948  & 0.999  & 0.948 & 0.83  & 0.612  \\
${\cal L}(0,2)$ & 0.952  & 1 & 0.952 &  0.889 & 0.665   & 0.967 & 0.999  & 0.965 & 0.993 & 0.588  \\
${\cal C}(0,2)$ & 0.974  & 1 & 0.974  &   & 0.537 & 0.983  & 1 & 0.981 &  & 0.526  \\ \hline
\end{tabular} 
}
\label{tableau1linear}
\end{table}

\begin{table}
\caption{Coverage probabilities for linear model, $\pi(X)=0.8$, $n=500,20$, error Gaussian, Laplace, Cauchy. }
\scriptsize{
\begin{tabular}{|c|ccccc|ccccc|} \hline
  &  &$n=$ & $100$&  & &  &$n=$ & $20$& &   \\
 $\varepsilon$    &  $CP_{LS}$ &  $CP_{LAD}$ & ${\hat {CP}_{LS}}$ &   $NCP_{LS}$ &  $NCP_{LAD}$ &$CP_{LS}$ &  $CP_{LAD}$ & ${\hat {CP}_{LS}}$   &  $NCP_{LS}$ &  $NCP_{LAD}$ \\ \hline
${\cal N}(0,1)$ &  0.954 & 0.948  & 0.951 & 0.964 & 0.800 &  0.961 &  0.913 & 0.943  & 0.947 & 0.848  \\
${\cal L}(0,1)$ &  0.950 & 0.985 &  0.944  & 0.962 &   0.811 & 0.965 & 0.958 & 0.952  & 0.956 & 0.806  \\
${\cal C}(0,1)$ &  0.983 & 1 &  0.978  &  & 0.769 &  0.992 & 0.999 & 0.978 &  & 0.656  \\ \hline
${\cal N}(0,2)$ & 0.952  & 0.999  & 0.950  & 0.949 & 0.561 & 0.968  & 0.997  & 0.946  & 0.870 & 0.630  \\
${\cal L}(0,2)$ & 0.955  & 1 & 0.952  & 0.958 & 0.583   & 0.964 & 0.998 & 0.946 & 0.993 & 0.620  \\
${\cal C}(0,2)$ &  0.982 & 1 & 0.982   &  & 0.54 & 0.985  & 1 & 0.976 &  & 0.481  \\ \hline
\end{tabular} 
}
\label{tableau2linear}
\end{table}

\begin{table}
\caption{Mean, standard deviation, median of parameter estimations for linear model, $\pi(X)=0.8+0.2|x-1|$, if $|x-1| \leq 1$ and 0.95 elsewhere, $\beta^0=10$, error Gaussian, Laplace, Cauchy. }
\scriptsize{
\begin{tabular}{|c|cc|cc|cc|} \hline
$n$ & mean & mean & sd & sd & median & median  \\
error & LS & LAD & LS & LAD &  LS & LAD  \\ \hline
$n=100$ &  &  &  &  &  &  \\ \hline
${\cal N}(0,1)$ & 10 & 10 & 0.07 & 0.09 & 10 & 10 \\
${\cal L}(0,1)$ & 10 & 10 & 0.10 & 0.08 & 10 & 10 \\
${\cal C}(0,1)$ & 9.94 & 10 & 93 & 0.12 & 9.98 & 10 \\
${\cal N}(0,2)$ & 10 & 10 & 0.15 & 0.18 & 10 & 10 \\
${\cal L}(0,2)$ & 10 & 10 & 0.21 & 0.16 & 10 & 10 \\
${\cal C}(0,2)$ & 10.7 & 10 & 59 & 0.25 & 9.92 & 10 \\ \hline
$n=20$ &  &  &  &  &  &  \\ \hline
${\cal N}(0,1)$ & 10 & 10 & 0.17 & 0.22 & 10 & 10 \\
${\cal L}(0,1)$ & 10 & 10 & 0.24 & 0.22 & 10 & 10 \\
${\cal C}(0,1)$ & 10 & 10 & 189 & 0.34 & 10 & 10 \\
${\cal N}(0,2)$ & 10 & 10 & 0.34 & 0.42 & 10 & 10 \\
${\cal L}(0,2)$ & 9.97 & 9.98 & 0.50 & 0.44 & 9.98 & 9.98 \\
${\cal C}(0,2)$ & 8.97 & 10 & 29 & 0.65 & 10 & 10 \\  \hline
\end{tabular} 
}
\label{tableau3linear}
\end{table}

\begin{table}
\caption{Mean, standard deviation, median of parameter estimations for linear model, $\pi(X)=0.8$, $\beta^0=10$, error Gaussian, Laplace, Cauchy.  True value $\beta^0=10$.}
\scriptsize{
\begin{tabular}{|c|cc|cc|cc|} \hline
$n$ & mean & mean & sd & sd & median & median  \\
error & LS & LAD & LS & LAD &  LS & LAD  \\ \hline
$n=100$ &  &  &  &  &  &  \\ \hline
${\cal N}(0,1)$ & 10 & 10 & 0.08 & 0.10 & 10 & 10 \\
${\cal L}(0,1)$ & 10 & 10 & 0.11 & 0.09 & 10 & 10 \\
${\cal C}(0,1)$ & 9.71 & 10 & 16 & 0.13 & 10 & 10 \\
${\cal N}(0,2)$ & 10 & 10 & 0.17 & 0.21 & 10 & 9.99 \\
${\cal L}(0,2)$ & 9.99 & 10 & 0.22 & 0.18 & 9.99 & 10 \\
${\cal C}(0,2)$ & 13 & 10 & 1.84 & 0.26 & 9.97 & 9.99 \\ \hline
$n=20$ &  &  &  &  &  &  \\ \hline
${\cal N}(0,1)$ & 10 & 10 & 0.20 & 0.23 & 10 & 10 \\
${\cal L}(0,1)$ & 10 & 10 & 0.26 & 0.25 & 10 & 10 \\
${\cal C}(0,1)$ & 8.46 & 10 & 31.2 & 0.38 & 9.99 & 9.99 \\
${\cal N}(0,2)$ & 9.99 & 9.98 & 0.37 & 0.47 & 9.98 & 9.98 \\
${\cal L}(0,2)$ & 10 & 10 & 0.53 & 0.50 & 10 & 10 \\
${\cal C}(0,2)$ & 9.22 & 10 & 34 & 0.82 & 10 & 10 \\ \hline
\end{tabular} 
}
\label{tableau4linear}
\end{table}

\begin{table}
\caption{The acceptance rate of hypothesis $H_0: \eb=\ebo$ with respect to  missing probability $1-\pi(X)$.} 
\scriptsize{
\begin{tabular}{|l|c|c|c|} \hline
  & $1-\pi(X)=0.20$ & $1-\pi(X)=0.50$ & $1-\pi(X)=0.80$ \\ \hline
complete data, LS method & 1 & 0.996 & 0.976 \\ 
complete data, LAD  method & 0.993 & 0.977 & 0.983 \\  
reconstituted data, LS method &  1 & 0.992 & 0.874 \\ \hline
\end{tabular} 
}
\label{tableau6}
\end{table}

\begin{table}
\caption{Empirical means, standard-deviations of differences between $Y_i$ and its reconstituted (forcasted) value,  with respect to estimation method and missing probability $1-\pi(X)$.} 
\scriptsize{
\begin{tabular}{|l|c|c|c|} \hline
  & $1-\pi(X)=0.20$ & $1-\pi(X)=0.50$ & $1-\pi(X)=0.80$ \\ \hline
mean($Y_i-\hat Y_{i,0,LS}$) & 0.066 & 0.066 &  0.066 \\ 
ean($Y_i-\hat Y_{i,0,LAD}$) & 0.321 & 0.321 &  0.321 \\ 
mean($Y_i-\hat Y_{i,LS}$) & 0.002 & -0.003 &  -0.0158 \\ \hline
sd($Y_i-\hat Y_{i,0,LS}$) & 3.35 & 3.35 &  3.35 \\ 
sd($Y_i-\hat Y_{i,0,LAD}$) & 3.36 & 3.36 &  3.36 \\ 
sd($Y_i-\hat Y_{i,LS}$) & 1.71 & 3.44 &  6.65 \\ \hline
\end{tabular} 
}
\label{tableau7}
\end{table}

\end{document}